# Enhancing Accessibility in Special Libraries: A Study on AI-Powered Assistive Technologies for Patrons with Disabilities

**Snehasish Paul[1*] Shivali Chauhan[2]**

[1]Librarian, Global Institute of Technology & Management, Gurugram, Haryana, India, 122506
**Email:** snehasishpaulas98@gmail.com

[2]Intern, Central Library, Jiwaji University, Gwalior, India, 474004
**Email:** chauhanshivali609@gmail.com

**ORCID- https://orcid.org/0009-0003-2730-5314**
***Corresponding author email:** snehasishpaulas98@gmail.com

## Abstract

This study seeks to identify the potential role of AI-driven assistive technologies in enhancing access to libraries for persons with varying degrees of challenges. Traditional libraries pose a problem to many users with vision and mobility, among other conditions related to physical and infirmities. This mixed-methods research approach will examine ways in which AI-powered assistive tools and applications associated with text-to-speech, navigation systems, and personalized assistants are revolutionizing library services through a literature review, survey methods, interviews, and case studies. Our findings suggest that these technologies greatly increase the autonomy and participation of people with physical disabilities, providing personalized support and access to a wide range of resources. From this, some key findings have been deduced from the research, showing a strong impact on user experience and efficiency in services, while at the same time bringing out important considerations related to privacy and ethical implementation. This study highlights the central role of AI in making library settings more inclusive, thereby allowing equal access to knowledge and participation in the community. Such insight thus serves professionals working in libraries, policymakers, and technology developers for innovations to occur uninterruptedly, with future research directions proposed that would refine such technologies, especially toward the special needs of diverse populations. By adopting AI, libraries could uphold their mission of providing equal access to knowledge through full and equal participation of all persons, regardless of any type of physical ability, in the learning and community activities carried out by the library. This study paves the way for future innovations in creating more accessible and inclusive library spaces.

## 1. Introduction

Libraries have remained relevant places for knowledge dissemination and community engagement since time immemorial. However, in conventional library spaces, there are many barriers to people who are physically challenged because of architectural barriers and service provision models that are not up to date. These had weighed down on the capacity of the visually challenged, mobility-impaired, or any other physically challenged individual using full library potential. AI is promising for solving these problems. Next-generation AI-driven assistive technologies, from advanced text-to-speech systems through AI-powered navigation aids and intelligent virtual assistants, will redefine library services in terms of their inclusiveness and user friendliness. This study attempts to explore the role that AI might currently or soon play in furthering library accessibility, based on a literature review supplemented by case studies and data analysis of the current experiences from the target communities on both the patron and professional sides. This study aims to assess the ways in which AI-powered tools contribute to physically challenged library users' independence and engagement by looking at the operational efficiency impact and exploring the adaptability of AI solutions to diverse needs. Furthermore, it highlights best practices and showcases possible future directions for integrating AI into library services. Libraries must learn how to effectively use AI technology to apply and surmount the particular challenges of accessibility. In this way, libraries provide better service to each member in their community by giving them equal access to resources and creating opportunities for learning and engaging. This reinforces the hope that current research findings will have far-reaching implications for library services, with a potential role for AI in enhancing accessibility across different public institutions in ways yet to be conceived of, revolutionizing how libraries can effectively address diverse populations.(Kirongo et al., 2022)

## 2. Objectives of the study

1. To evaluate AI-driven assistive technologies to improve library accessibility for mobility-impaired patrons.

2. To assess the impact of AI interfaces on the independence and user experience of physically challenged library users.

3. To analyze AI's potential to enhance service efficiency for patrons with mobility impairments.

4. To explore AI adaptability in meeting the diverse needs of physically challenged library users.

5. To investigate AI's role in creating inclusive library environments in the Web 3.0 era.

## 3. Literature review

There are only a limited number of published articles on this topic. In this study, the following papers were reviewed.

Kirongo et al. (2022) studied three pilot schools and an assistive technology center hosted at the Meru University of Science and Technology. Their research focused on broadening the understanding of learners with physical disabilities and innovators to support appropriate identification, assessment, and selection of relevant AI-ATs. This study established a critical call for distinctive technological propositions to meet various needs and highlighted the role of artificial intelligence in adaptive learning environments. Adel et al. (2023) developed an intelligent robot assistant that combines technologies such as AI, ML, and IoT. Such technologies allow robots to integrate many devices, provide navigation assistance, and automate tasks relevant to and reassuring the performance of enhancements that comprise the daily life activities of the visually impaired. AI and ML can enable a robot to learn from and adapt to user preferences and the environment in which it operates, making it more valuable and powerful. Herold et al. (2022) conducted a psychology-based assessment of BERT's representation of distality, using a Stereotype Content Model. This study revealed a bias along established stereotype dimensions in BERT, mainly affecting perceptions of warmth and competence among people with disabilities. This bias shows some systemic under- or misrepresentation of disabilities, either in the training data or in the model architecture of BERT, which then influences its output and applications. Gujral et al. (2020) elaborated on ways in which an AI-based assistive brings a sea of change in the organizational setup within the precincts of academic libraries, including text-to-speech conversion, image recognition for document digitization, and adaptive interfaith users who have disabilities or language barriers. Critical among these developments, despite challenges such as ensuring privacy and mitigating bias, AI will continue to be important in creating fairer learning environments through academic libraries. Bi et al. (2022) underlined the huge transformative potential of associated technologies in shaping the future of intelligent libraries, where such an environment would be adaptive in allocating resources for maximal user experience and lifelong learning initiatives. They provided an overview of AI- and IoT-based technologies in brilliant libraries, in terms of applications for innovative services, sustainability, and security. This shows that such technologies improve the accessibility of libraries to physically disabled patrons through personal services and assistive technologies, in which intelligent navigation systems and adaptive interfaces occupy a central place. Artificial intelligence, as embedded in the model, facilitates personalization through recommendations and real-time assistive solutions, where IoT supports sustainability initiatives, energy management, and environmental monitoring.

## 4. Overview of Assistive Technology and AI-Based Assistive Technology

Assistive technologies allow libraries to provide access and emphasize their efforts to serve disabled patrons. Adjustable workstations and furniture make working easier when users with physical disabilities access the library resources. Alternative input devices, such as alternative keyboards and mice, assist users who may have problems using the so-called 'standard' equipment. Magnification and text-to-speech tools, such as CCTVs and magnifiers, screen readers for those with visual difficulties or reading problems. Libraries also help users obtain materials in alternative formats, such as braille, large prints, or audio, to help them interact with materials for people with a wide range of disabilities. Other learning aids and productivity enablers include assistive software such as screen readers, voice dictation, and mind-mapping tools (Tripathi & Shukla, 2014).

AI-powered assistive technologies can change everything about library accessibility for people with mobility impairment. In other words, AI can enable libraries to be more inclusive, offer equal opportunities for all, and allow everyone to use available information and resources. In most instances, AI-based assistive technologies are revolutionizing accessibility to libraries for many people with mobility impairments. They apply artificial intelligence to ensure that they offer navigation assistance based on customer needs, intuitive interfaces, and better resource discovery. Key features often include voice control, computer vision for identifying books and obstacles, and predictive algorithms for anticipating user needs. AI assistants, after the library management systems are integrated, personalize book recommendations to each user and guide them to the materials of interest. Furthermore, it enables access to the material digitally. The technology puts forth equalization by giving users independence, improving the user experience in libraries, and promoting inclusiveness within the cultural space of Web 3.0 and beyond. (LIM, 2023)

| Product/Service | Feature 1 | Feature 2 | Feature 3 | Feature 4 |
|---|---|---|---|---|
| Voiceitt | Real-time speech translation | Customizable | Device integration | Supports languages |
| NoorCam MyEye | Text recognition | Face and product identification | Colour and money recognition | Portable design |
| Braina AI | Voice recognition | Text-to-speech | Task management | |
| Ava | Real-time transcription | Speaker ID | Multilingual | |
| Sesame Enable | Hands-free device control | Gesture commands | | |
| Proloquo2Go | Symbol-based communication | Customizable vocabulary | | |
| SpectrumNews.org | Autism research | Neurodevelopment | | |
| Envision AI | Text-to-speech | Scene description | Barcode, colour recognition | |

**Table 1:** Comparison of AI-Driven Assistive Technologies and Their Features

This table compares the various AI-based assistive technologies with respect to their features. It presents products such as Voiceitt, NoorCam MyEye, and EnVision AI, including speech translation, text recognition, and hands-free control. This overview is critical in understanding the various technologies that could improve accessibility within libraries, specifically for people with mobility impairments, by providing varied solutions for communication, information access, and interaction within the environment (*GeeksforGeeks*, 2024).

| AI-Based Assistive Technology | Features |
|---|---|
| **AI-Powered Optical Character Recognition (OCR) Systems** | - Accurately extracts and interprets text from scanned pictures and documents.<br>- Allows visually impaired people to access a variety of reading resources. |
| **Text-to-Speech Conversion for Audiobooks** | - Converts printed materials into high-quality audiobooks<br>- Leverages natural language processing and machine learning algorithms |
| **Natural Language Processing (NLP) for Voice-Based Searches** | - Enables voice-based searches within library catalogues and databases<br>- Allows visually impaired individuals to search using voice commands |
| **AI-Powered Image Recognition and Description** | - Automatically analyses and describes visual content in books, documents, and digital resources.<br>- Provides detailed descriptions of images, diagrams, and other visual elements.<br>- Enables visually impaired users to better understand and appreciate image-rich library materials. |
| **AI-Driven Accessibility Evaluation Tools** | - Scans websites, documents, and digital content for accessibility issues<br>- Identifies problems such as missing alt-text, low colour contrast, or lack of keyboard navigation<br>- Helps libraries create more inclusive and accessible digital environments |
| **AI-Powered Personalized Recommendations** | - Analyses a visually impaired patron's reading history, preferences, and accessibility needs<br>- Provides personalized book recommendations and resource suggestions<br>- Helps visually impaired users discover new materials tailored to their interests and accessible formats |
| **AI-Driven Accessibility Monitoring and** | - Tracks the usage and effectiveness of assistive technologies and accessibility features |
| **Feedback** | - Analyses usage data and user feedback to continuously improve accessibility offerings<br>- Provides instant feedback and guidance to visually impaired users via chatbots and virtual assistants |
| **AI-Powered Accessibility Compliance and Auditing** | - Helps libraries ensure compliance with accessibility standards and guidelines (e.g., WCAG, ADA)<br>- Automates accessibility audits and compliance checks<br>- Identifies and addresses accessibility issues efficiently to ensure digital resources and services are accessible to all users |

**Table 2:** AI-Powered Assistive Technologies for Visually Impaired Library Patrons

The table provides an overall description of AI-driven assistive technologies that could improve accessibility for visually impaired library patrons, including AI-powered OCR systems, text-to-speech conversion for audiobooks, natural language processing for voice-based searches, AI-powered image recognition and description, AI-driven accessibility evaluation tools, AI-powered personalized recommendations, AI-driven accessibility monitoring and feedback, and AI-powered accessibility compliance and auditing. The table for each of these technologies enumerates the key features and associated benefits that illustrate how AI-driven solutions can significantly improve the library experience of visually impaired users, in terms of better access to information, better navigation across resources, and more personalized assistance (*GeeksforGeeks, 2024)*.

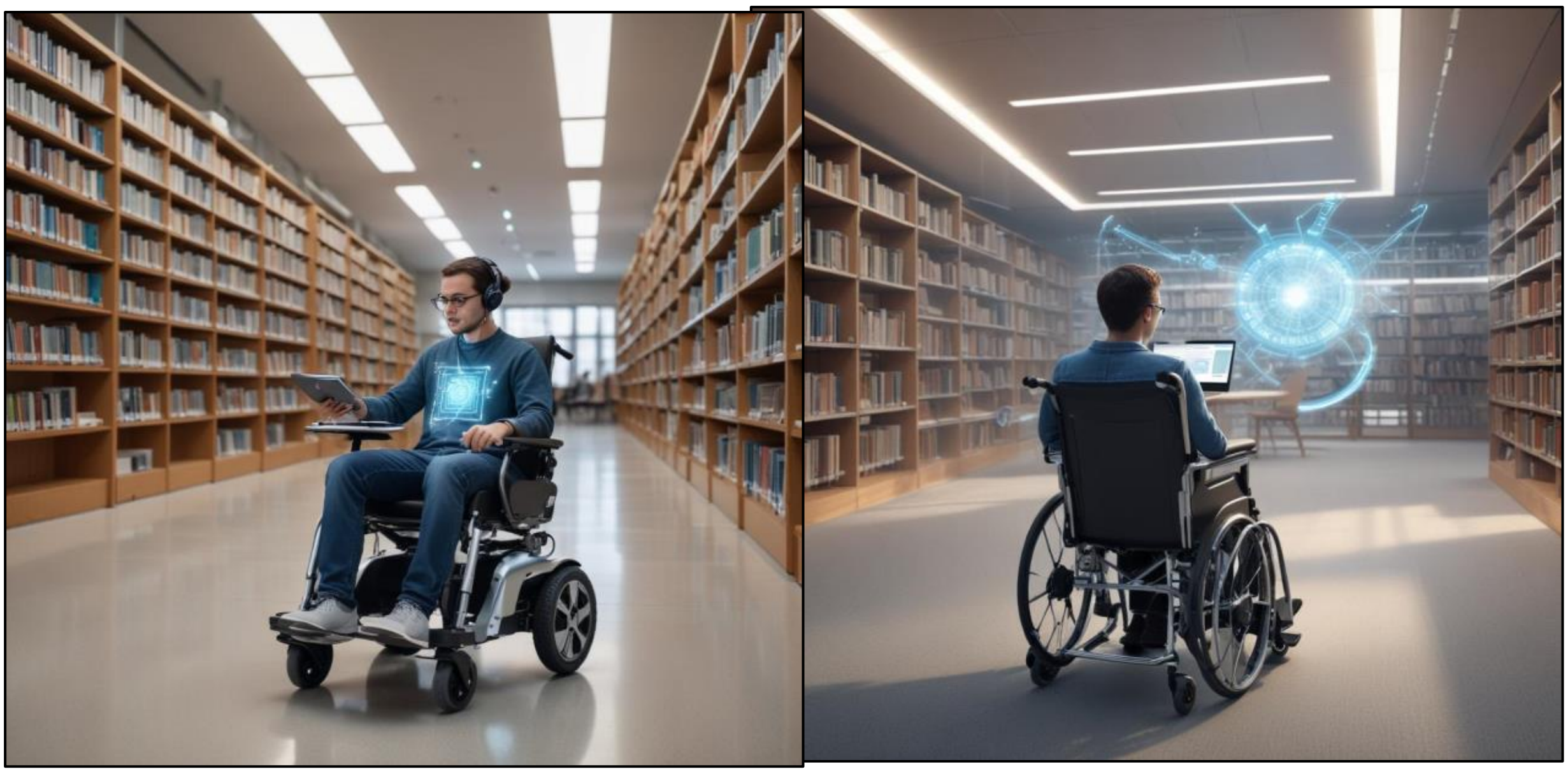

**Figure 1.** AI-assisted library navigation and resource access for wheelchair users.

**Image source (FREEP! K AI generator)**

AI-powered assistive technologies have significantly affected library access for people with mobility impairment. It initiates independent navigation and resource use through wheelchair-friendly designs, personalized AI interfaces, and digital access tools. In a sense, this technology links traditional library environments with modern AI solutions and creates inclusive places where knowledge expands toward every person, regardless of their physical abilities (*Freepik AI Image Generator* - *Free Text-to-Image Generator*, 2024).

## 5. Methodology

The methodology used in this study on the accessibility of AI-empowered libraries was a rigorous mixed-methods approach. First, a literature review helps to establish a point on the existing knowledge base regarding AI assistive technologies and accessibility to libraries. Quantitative data were gathered from questionnaires sent to five special libraries, and questionnaires were shared with 20 library patrons with mobility impairment. This provides a general overview of the existing practices and user needs. This was complemented by qualitative research, including semi-structured interviews with four librarians and five accessibility experts, as well as focus groups involving 30 patrons, providing in-depth insight into experiences and challenges. Four detailed case studies of libraries that successfully used AI technologies provided concrete examples of the best practices. conducted a technology assessment rating of the current tool in terms of usability, with the assistance of 20 participants with different mobility impairments. It includes statistical methods for quantitative data, on the other hand, and thematic content data analysis for qualitative information, hence securing a rich, multifaceted understanding of the subject. In the course of this research, we would strongly emphasize ethical considerations: securing the necessary approval and ensuring the confidentiality of the participants. This rigorous methodology will help us delve deeply into how AI-driven assistive technologies are entirely changing library accessibility for people with mobility impairments and provide valuable lessons for future rollouts and improvements in this vital area.

## 6. Data Collection Method

A questionnaire was administered to 20 physically challenged clients through Google Forms to retrieve their experiences and needs regarding the AI-driven assistive technologies offered in libraries. The descriptive data obtained showed that many of these people access services daily and provide high ratings of accessibility, thus showing satisfaction, although some have pointed out areas for improvement. Although an overwhelming majority of the participants confirmed a rather clear latent need for interactive and real-time AI technologies, very few gave concrete suggestions, indicating a distinct need to raise awareness of what aids AI can offer. This means that in addition to providing services to these patrons with special mobility-impaired needs, libraries will also have to create AI tools to aid step one of instantaneous responses and then conduct more in-depth research in measuring those needs more appropriately, with education programs designed to inform about these technologies.

### 6.1 Data Analysis Procedures

The data collected from this research were analysed using Excel for quantitative insights and NVivo for qualitative analysis to provide a comprehensive overview that represents the experiences and needs of users.

| Participant Code | Age Group | Gender | Disability Type | AI Integration Suggestion |
|---|---|---|---|---|
| P1 | 35-45 | Female | Visual Impairment | Implement AI-powered text-to-speech systems for audiobooks. |
| P2 | 20-30 | Male | Mobility Impairment (wheelchair user) | Introduce AI-driven navigation assistance for easy library access. |
| P3 | 50-60 | Male | Hearing Impairment | Deploy AI-based real-time captioning for library events. |
| P4 | 18-25 | Female | Cognitive Impairment (dementia) | Utilize AI for personalized memory aids and reminders. |
| P5 | 30-40 | Male | Neurological Disorder (Parkinson's) | Integrate AI for voice-controlled library catalogue searches. |
| P6 | 55-65 | Female | Psychiatric Disability (depression) | Develop AI-powered mood-tracking apps for personalized reading recommendations. |
| P7 | Oct-18 | Male | Developmental Disability (autism) | Use AI for personalized learning tools and sensory-friendly library spaces. |
| P8 | 25-35 | Female | Speech and Language Disorder | Implement AI speech recognition for interactive library services. |
| P9 | 40-50 | Female | Chronic Health Condition (diabetes) | Utilize AI for health monitoring and personalized wellness programs. |
| P10 | 15-25 | Male | Learning Disability (dyslexia) | Introduce AI-powered reading assistance tools and adaptive learning resources. |
| P11 | 28-38 | Female | Visual Impairment | Implement AI-driven image recognition for accessing visual content. |
| P12 | 22-32 | Male | Mobility Impairment (cane user) | Deploy AI for predictive maintenance of accessibility equipment. |
| P13 | 58-68 | Female | Hearing Impairment | Introduce AI-powered sign language translation services. |
| P14 | 20-30 | Male | Cognitive Impairment (intellectual disability) | Use AI for personalized cognitive skill development programs. |
| P15 | 32-42 | Female | Neurological Disorder (multiple sclerosis) | Implement AI for adaptive seating and environment control systems. |
| P16 | 60-70 | Male | Psychiatric Disability (schizophrenia) | Develop AI-driven social interaction aids and virtual support groups. |
| P17 | 20-30 | Female | Developmental Disability (Down syndrome) | Utilize AI for adaptive learning tools and social skills development. |
| P18 | 30-40 | Male | Speech and Language Disorder | Implement AI for personalized speech therapy exercises. |
| P19 | 45-55 | Female | Chronic Health Condition (chronic pain) | Use AI for pain management and adaptive seating recommendations. |
| P20 | 18-28 | Male | Learning Disability (dyscalculia) | Introduce AI-powered math tutoring and adaptive learning tools. |

**Table 3:** Participant Characteristics and AI Integration

This table includes 20 participants, from P1 to P20, of varying ages and genders, who have different kinds of disabilities, such as visual, mobility, hearing, and cognitive impairments. Each participant was provided AI-based integration recommendations for special libraries, including text-to-speech systems, navigation aids, real-time captions, memory aids, and adaptive learning tools. All these facilities help make libraries more accessible and better support their physically challenged patrons, thereby promoting inclusiveness in accessing library resources.

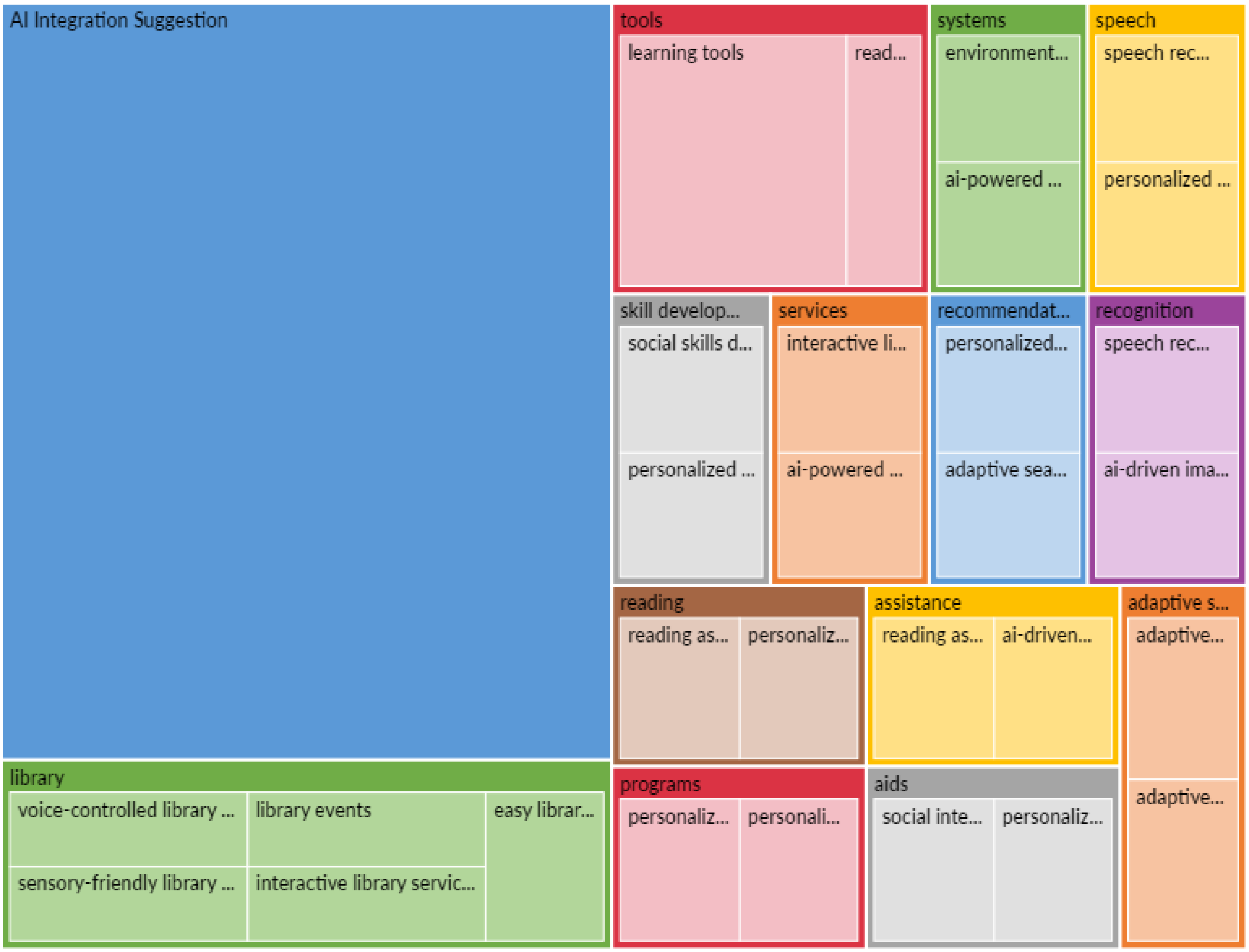

**Figure 2:** Hierarchy Chart of AI Integration Suggestions for Assistive Technologies

A hierarchy chart of AI integration suggestions for assistive technologies was generated in NVivo. The plot provides a clear overview of some critical areas to focus on when designing support for people with visual impairments: tools for learning, reading, AI-powered and individualized environmental systems, speech recognition technologies, and skill development programs, in particular, enhancing social skills. It also displays several services, including interactive services, AI-driven assistance, recommenders, and adaptive searching. It showcases reading aids, and there is a palette of accessible library services, from voice-controlled to sensory-friendly. Another component common to all categories is the concept of personalization, which reinforces the key theme of tailoring technologies to the requirements of the individual. The dominance of AI integration has the potential to heighten the adaptability and effectiveness of assistive technologies. It spans from basic reading aids to complex social skills development, opening up a large number of directions for research and development in assistive technologies for AI-enhanced visual impairment.

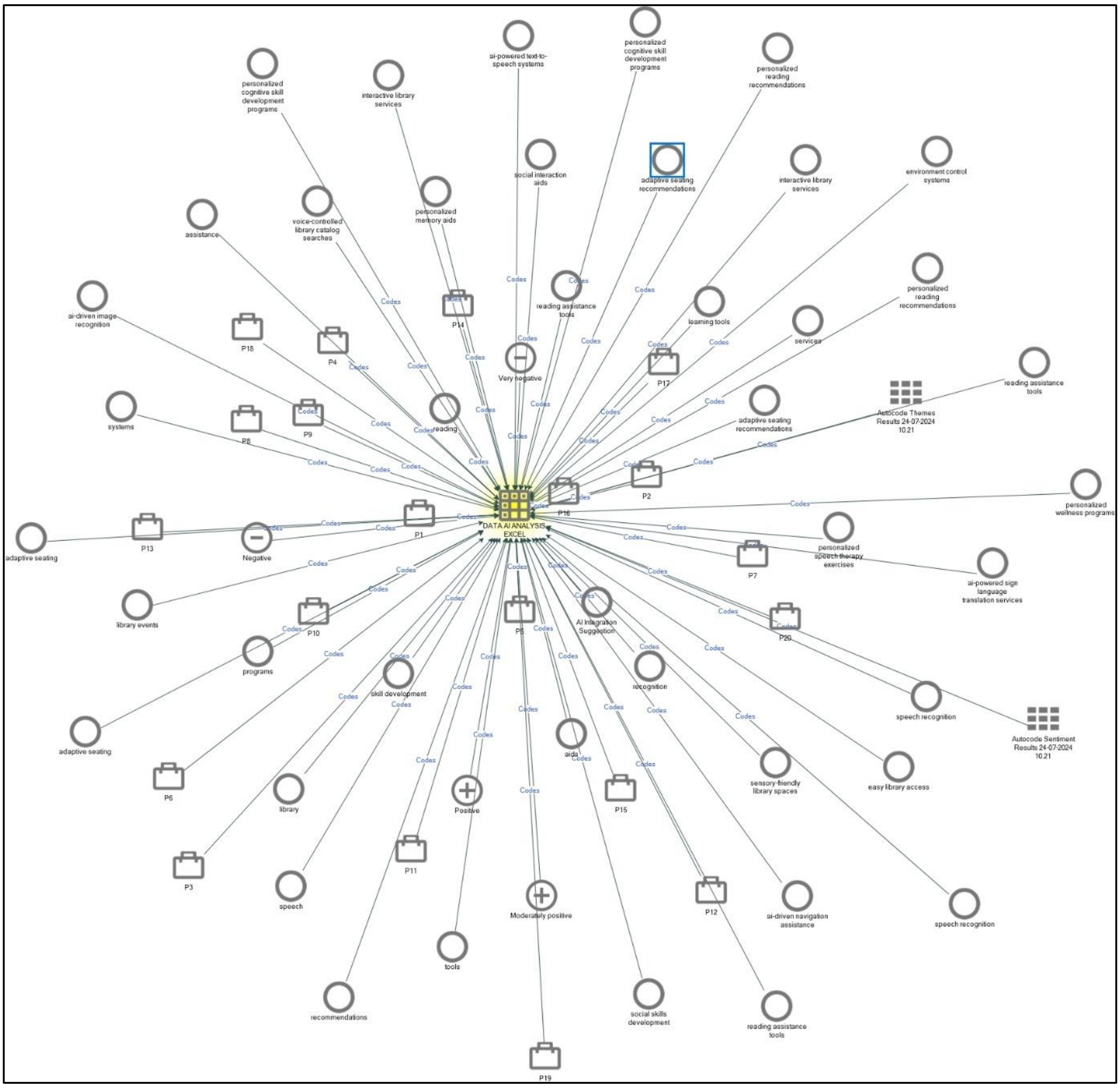

**Figure 3: Explore Diagram of Survey Data Created with NVivo**

This exploration diagram, generated in NVivo, visualizes survey responses to assistive technologies for visual impairment via Google Forms. The central concept, represented by a node, branches into other interconnected nodes representing devices, features, and functionalities. Basic themes include accessibility, personalization, and sensory augmentation. The structure is radial, showing how these very different aspects finally support the visually impaired and giving an overview of the whole spectrum of responses to the survey, emphasizing multi-facetedness.

## 7. Case Studies:

Some successful case studies on AI-based assistive technologies implemented in libraries to pave the way for accessibility to patrons with a disability profile:

### 7.1 New Jersey State Library's Talking Book and Braille Center

The Talking Book and Braille Center is a program in the New Jersey State Library that operates text-to-speech platforms to convert printed text into audio formats for visually impaired patrons. Moreover, it offers training in the use of assistive technologies, such as mobile devices, focusing on the library's mission as a learning center, and providing access to information for all (Team, 2017).

### 7.2 University of North Texas Libraries

The University of North Texas Libraries has integrated an AI chatbot named "Ask Us" for 24/7 assistance with patrons. This chatbot assists with various questions, from directional to complex research needs. It was undertaken as a way to improve user satisfaction and engagement by instantly answering the most asked questions so librarians might handle long, in-depth research support (Sanji et al., 2022).

### 7.3 University of Lagos

The University of Lagos has introduced AI in some of its library services, a milestone in Nigeria's academic library landscape. This move, which involves the adoption of AI technologies, aims to improve service delivery and enhance user satisfaction using quick access to information and smoothing library operations. It finally underlines the fact that AI, despite existing challenges in developing countries, can transform library services (Moustapha, 2023).

## 8. Collection Development and Data Analytics

AI-powered tools have been harnessed in collection development to aid informed decisions regarding library materials. Librarians also offer insights into diversification and make accessible diverse collections through AI. It also helps libraries gain insight into user behavior through data analytics, which AI is capable of, to optimize services accordingly to suit changing community needs.(Woldetsadik et al., 2024)

Library AI services around the world have changed how they work and the services provided to their patrons. By improving information retrieval for users' supply with personalized assistance, increasing accessibility, and digital preservation, AI has a real transformative effect in this library landscape. As libraries continue to adopt such technologies, they need to balance innovation with ethical concerns and user privacy.

## 9. Conclusion

AI-powered assistive technologies are the most significant development in terms of accessibility and inclusivity in libraries. AI assistive technologies would provide access to a wide array of resources and allow more independence for visitors with physical impairments to libraries. From smart devices with text-to-speech features to AI-driven chatbots, enabling real-time information and support, these technologies enhance user experience and provide equal opportunities for all community members. However, security, reliability, and user privacy concerns must be considered with utmost sincerity when implementing AI-driven assistive technologies. In addition to the overall training of staff and patrons, measures against such challenges must be taken so that libraries continue to be trusted and accessible spaces for all. With AI constantly improving in the future, we can look forward to further innovations that continue to stretch the envelope of library accessibility. Keeping pace with and upholding such innovations, libraries have become beacons for knowledge, learning, and community engagement in the new information age. In turn, the continuous development and integration of AI technologies increasingly bind the future of the library as an inclusive and accessible space (Onyango et al., 2023).

## 10. Future research:

Future directions of AI-driven assistive technologies for improving access to books in Indian libraries vary, from effectiveness evaluation and cultural adaptation to system integration, ethical considerations, and socioeconomic impact. This corresponds to testing user satisfaction, personalization of technology concerning India's linguistic diversity, privacy concerns, and an overall long-term sustainability assessment. As such, this research will be imperative for fine-tuning technologies, shaping inclusive library services in India, and potentially influencing global practices to improve access to books for patrons with mobility impairments (Dange et al., 2023).

## 11. Recommendations for Implementation of AI-based Assistive Technologies in Indian Libraries

Although AI-driven assistive technologies for book access have not yet been integrated into Indian libraries, their successful implementation elsewhere opens up an avenue. Guided by Ranganathan's Third Law of Library Science, **"Every book its reader,"** we can leverage technology to ensure that every reader, regardless of mobility impairments, has access to books. In this regard, the following recommendations are proposed for Indian libraries (Ranganathan, 1931).

### 11.1 AI-powered book retrieval systems:

1. Implement AI-driven robotic systems for the retrieval of physical books for patrons with mobility impairments.
2. Come up with voice-activated commands to request books.

### 11.2 Enhanced Digital Book Access:

1. Use an AI-powered OCR to digitally unlock access to physical books.
2. Integrate Text-to-Speech technology to produce an audio version of a book in the library collection.

### 11.3 Personalized Book Recommendations:

1. Run AI algorithms that suggest books on a patron's interests and accessibility needs.
2. Design systems that consider constraints on patron mobility in their suggestions for books.

### 11.4 AI-Assisted Navigation:

Install AI-powered indoor navigation systems that can guide especially abled patrons with mobility impairments to their required sections in books.

### 11.5 Adaptive Reading Devices:

Introduce AI-enhanced e-readers that adapt to different uses by providing page-turning mechanisms for people who can only slightly move their hands.

**11.6 Virtual library assistance:**

Develop AI chatbots to assist patrons in finding and obtaining books and answering questions about book availability and accessibility.

**11.7 Remote Book Access:**

Utilize AI to manage an effective system of home delivery or digital access to books for users who cannot visit the library.

**11.8 Staff Training:**

Develop programs to train library staff to use such technologies in an approachable and efficient manner.

**11.9 Collaborative book-sharing:**

Integrating AI-driven interlibrary loan systems to open access to locally unavailable books

**11.10 Continuous Improvement:**

Use AI to examine use patterns and obtain feedback to improve access to books for physically challenged patrons. Adopting AI-driven solutions in Indian libraries could increase accessibility to books for patrons with mobility-related disabilities; hence, physical limitations become no obstacle to saving one's quest for knowledge and reading. Such an approach would be in tune with Ranganathan's principles and would place the Indian libraries' threshold of playing a pioneering role in inclusive access to books (A. & Gozali, 2024).

## Conflict of interest

The authors declare that they have no competing interests.